\documentclass[
	aps, prb,
	superscriptaddress,
	epsf,
	amsmath, amssymbol,
	twocolumn,
	]{revtex4-1}

\usepackage[dvipdfmx]{graphicx}
\def\deg{\ensuremath{^{\circ}}}
\usepackage[dvipdfmx]{color}
\ifx\color@rgb\@undefined\else
  \definecolor{kw1}{rgb}{0,0,1}
  \definecolor{kw2}{rgb}{1,0,0.4}
  \definecolor{kw3}{rgb}{0,1,0.4}
  
\fi

\begin{document}


\title{Magnetism behavior of $T^{\prime}$-type Eu$_2$CuO$_4$ revealed by muon spin rotation/relaxation measurements}

\author{M. Fujita}
	\thanks{fujita@imr.tohoku.ac.jp}
	\affiliation{Institute for Materials Research, Tohoku University, 2-1-1 Katahira, Aoba-ku, Sendai 980-8577, Japan}
\author{K. M. Suzuki}
	\affiliation{Institute for Materials Research, Tohoku University, 2-1-1 Katahira, Aoba-ku, Sendai 980-8577, Japan}
\author{S. Asano}
	\affiliation{Department of Physics, Graduate School of Science, Tohoku University, 6-3 Aoba, Aramaki, Aoba-ku, Sendai 980-8578, Japan}
\author{H. Okabe}
	\affiliation{Muon Science Laboratory, Institute of Materials Structure Science, High Energy Accelerator Research Organization (KEK), Tsukuba, Ibaraki 305-0801, Japan}
\author{A. Koda}
	\affiliation{Muon Science Laboratory, Institute of Materials Structure Science, High Energy Accelerator Research Organization (KEK), Tsukuba, Ibaraki 305-0801, Japan}
\author{R. Kadono}
	\affiliation{Muon Science Laboratory, Institute of Materials Structure Science, High Energy Accelerator Research Organization (KEK), Tsukuba, Ibaraki 305-0801, Japan}
\author{I. Watanabe}
	\affiliation{Advanced Meson Science Laboratory, Nishina Center for Accelerator-Based Science, The Institute of Physical and Chemical Research (RIKEN), Wako, Saitama 351-0198, Japan}
\date{\today}

\begin{abstract}

We performed muon spin rotation/relaxation measurements to investigate the magnetic behavior of $T^{\prime}$-type Eu$_2$CuO$_4$ (ECO), which is the parent compound of electron-doped cuprate superconductors, and the effects of oxygen-reduction annealing on its magnetism. In as-sintered (AS) ECO, we clarified the development of magnetic correlations upon cooling below $T_\mathrm{N1}$ (= 265 K) as well as the coexistence of a dominant fluctuating spin state and partially ordered spin state in the temperature range between $\sim$150 K and $T_\mathrm{N1}$. Upon further cooling, uniform long-range magnetic order was observed below $T_\mathrm{N2} = 110$ K, which is close to the ordering temperature of 115 K in $T^{\prime}$-type La$_2$CuO$_4$ (LCO) [Phys. Rev. B {\bf 82}, 180508(R) (2010)]. For oxygen-reduction-annealed ECO, a similar ordering sequence with the same $T_\mathrm{N2}$ was observed but without the partially ordered spin state. Therefore, the fluctuating spin state over a wide temperature range and a $T_\mathrm{N2}$ less than the N{\'e}el temperature ($T_\mathrm{N2} \approx T_\mathrm{N1}$) in $T$-type LCO are common features of the $T^{\prime}$-type parent $R_2$CuO$_4$ ($R$CO, $R$: rare-earth ion). The origin of the partially ordered spin state in AS ECO is discussed from the viewpoint of chemical defect. Furthermore, we discuss the roles of electron doping and repairing defect in the observed effect of annealing on the magnetism of $T^{\prime}$-type $R$CO. 

\end{abstract}


\maketitle

\section{Introduction}

The discovery of superconductivity in $T^{\prime}$-type $R_2$CuO$_4$ ($R$CO, $R$: rare-earth) without Ce substitution stimulated research on the ground state of host materials in terms of the oxygen coordination around Cu~\cite{Tsukada2005, Matsumoto2009, Asai2011, Takamatsu2012}. It is well known that high-temperature-synthesized standard samples of $T^{\prime}$-type $R$CO with square planar coordination exhibits N{\'e}el order at $\sim$250 K\cite{Luke1990}, similar to the case of $T$-type La$_2$CuO$_4$ (LCO) with six-oxygen coordination~\cite{Uemura1989}. However, Hord and {\it et al}. reported through muon spin rotation/relaxation ($\mu$SR) measurements a lower ordering temperature in non-superconducting $T^{\prime}$-type LCO\cite{Hord2010}, which is synthesized using a soft chemical method. Spin correlations start to develop upon cooling below $T_\mathrm{N1} \approx 200$ K, and a long-range magnetic order emerges at $T_\mathrm{N2} = 115$ K, showing a successive ordering sequence. 
It is considered that the lower $T_\mathrm{N2}$ is due to enhanced two-dimensionality and related to the real ground state, namely, the metallic state of the ideal $T^{\prime}$-type $R$CO, which is suggested by previous theoretical studies~\cite{Das2009a, Weber2010a, Weber2010b}. 

Subsequently, through $\mu$SR measurements, Kojima {\it et al}. ~\cite{Kojima2014} reported the absence of long-range magnetic order in thin films of superconducting (SC) $T^{\prime}$-type $R$CO. The reason for the distinct ground states in different samples is understood in terms of chemical defect~\cite{Krockenberger2013, Naito2016}: Defects in a sample induce magnetic order due to strong electron correlations, but no such defects yield the real ground state of $T^{\prime}$-type $R$CO. Furthermore, it is considered that both as-prepared powder samples and thin films contain defects such as excess oxygen at the apical positions~\cite{Radaelli1994c, Schultz1996} and copper-ion deficiency~\cite{Kang2007}. For thin films, the defects can be effectively removed through annealing because of their large surface-to-volume ratio. In contrast, the complete removal of defects  from an entire sample is difficult for bulk compounds~\cite{Naito2016, Adachi2013}. In this context, in the low-temperature-synthesized LCO, partial defect induces magnetic order with a lower $T_\mathrm{N2}$. The ordered phase vanishes when the complete removal of the defects realizes a uniform electrostatic potential on the CuO$_2$ plane. 
To support the above scenario, a suppression of magnetic order accompanied by the appearance of superconductivity was reported by annealing the low-temperature-synthesized $T^{\prime}$-type La$_{1.8}$Eu$_{0.2}$CuO$_4$ (LECO) ~\cite{Adachi2016, Fukazawa2017}. 
However, many experiments, including recent X-ray absorption fine-structure measurements, have indicated electron doping as an alternative aspect of annealing treatment~\cite{Jiang1993, Higgins2006, Asano2018, Asano2020}. Therefore, the common magnetism in $T^{\prime}$-type $R$CO and the effect of annealing on the physical properties are fundamental issues in the research on superconductivity in $T^{\prime}$-type cuprates. 

To address these issues, a comprehensive investigation of high-temperature-synthesized non-SC bulk samples of $T^{\prime}$-type $R$CO is necessary, but the precise thermal evolution of magnetism and the effect of annealing on magnetism have not yet been studied for the host compounds. Hence, we performed $\mu$SR measurements on as-sintered (AS) and annealed (AN) $T^{\prime}$-type Eu$_2$CuO$_4$ (ECO). Owing to the absence of a rare-earth magnetic moment in ECO\cite{Tovar1989, Tanikawa1994}, we can extract the inherent magnetism of the CuO$_2$ plane in the $T^{\prime}$-type parent cuprates. We found that both AS and AN ECO have the same magnetic evolution as $T^{\prime}$-type LCO. Thus, the sequential ordering process with a lower $T_\mathrm{N2}$ is a common feature of magnetism in $T^{\prime}$-type $R_2$CuO$_4$. 
In addition, in the AS sample, we found the coexistence of a fluctuating spin state and partially ordered spin state in the temperature range of $\sim$150 K ($>$ $T_\mathrm{N1}$) to $T_\mathrm{N2}$. The partially ordered state disappears with annealing, indicating that the coexistence of the two states originates from defects existing in AS ECO. 


The remainder of this paper is organized as follows. Section II describes the sample preparation and experimental details. Section III presents the results of zero-field (ZF) and longitudinal-field (LF) $\mu$SR measurements. Section IV discusses the common features of the ordering sequence and the effect of annealing on the magnetism of $T^{\prime}$-type $R$CO. Finally, Sec. V summarizes this paper.

\section{Experimental details}

We prepared polycrystalline samples of ECO by using a standard solid-state reaction method. A pre-fired powder was pressed into pellets and sintered in air at 1050\deg C. We estimated the amounts of removed oxygen ($\delta$) to be $\approx 0.02$ with an accuracy of 0.001 based on the weight loss of the samples through the reduction annealing process. Furthermore, we confirmed the phase purity of the samples by x-ray powder diffraction measurements and determined the lattice constants from the diffraction pattern. The evaluated lattice constants with $I4/mmm$ crystal symmetry are summarized in Table \ref{Table}. None of our samples showed superconductivity, even after the annealing procedure. 

\begin{table}[b]
	\caption{Lattice constants and characterized magnetic ordering temperatures $T_\mathrm{N1}$ and $T_\mathrm{N2}$ for as-sintered (AS) and annealed (AN) Eu$_2$CuO$_4$.}
	\label{table_latticeparameter}
	\begin{tabular}{llllll}
		\hline
		&&a (\AA)&c (\AA)&$T_\mathrm{N1}$ (K) & $T_\mathrm{N2}$ (K)\\
		\hline
		Eu$_2$CuO$_4$&AS & 3.9049(1) & 11.9121(3) & 265(10) & 110(10)\\
		&AN & 3.9047(1) & 11.9126(3) & 250(10) & 110(10)\\
		\hline
	\end{tabular}
	\label{Table}
\end{table}

ZF and LF $\mu$SR measurements were performed at the Materials and Life Science Facility (MLF) at J-PARC, Japan and at the RIKEN-RAL Muon Facility of the Rutherford Appleton Laboratory (RAL), UK using single-pulsed positive surface muon beams. The samples were cooled to 10 K using an open-cycle $^4$He-flow cryostat. The $\mu$SR time spectrum, i.e., the time evolution of the asymmetry of positron events, is given by $A(t) = [F(t) - \alpha B(t)]/[F(t) + \alpha B(t)]$, where $F(t)$ and $B(t)$ are histograms of the positron event at the detectors positioned upstream and downstream of the sample, respectively, along the muon momentum direction. Further, $\alpha$ is a calibration factor for the efficiency and solid angle of the detectors. We evaluated $\alpha$ from an oscillation spectrum under a weak transverse magnetic field at approximately 300 K and calibrated the spectra for each sample. The obtained $\mu$SR time spectra were analyzed using the WiMDA program~\cite{Pratt2000}. 

Following previous studies~\cite{Kubo2002, Fujita2003}, we define $T_\mathrm{N1}$ ($T_\mathrm{N2}$) as the temperature at which the ZF $\mu$SR spectra changes from Gaussian-type decay to exponential-type decay (the temperature at which the bulk oscillation component appears) upon cooling. Table \ref{table_latticeparameter} summarizes $T_\mathrm{N1}$ and $T_\mathrm{N2}$.

\section{Results}\label{Results}

\subsection{Zero-field $\mu$SR measurements}\label{ZF}

\begin{figure}[t]
	\includegraphics[width=0.72\linewidth]{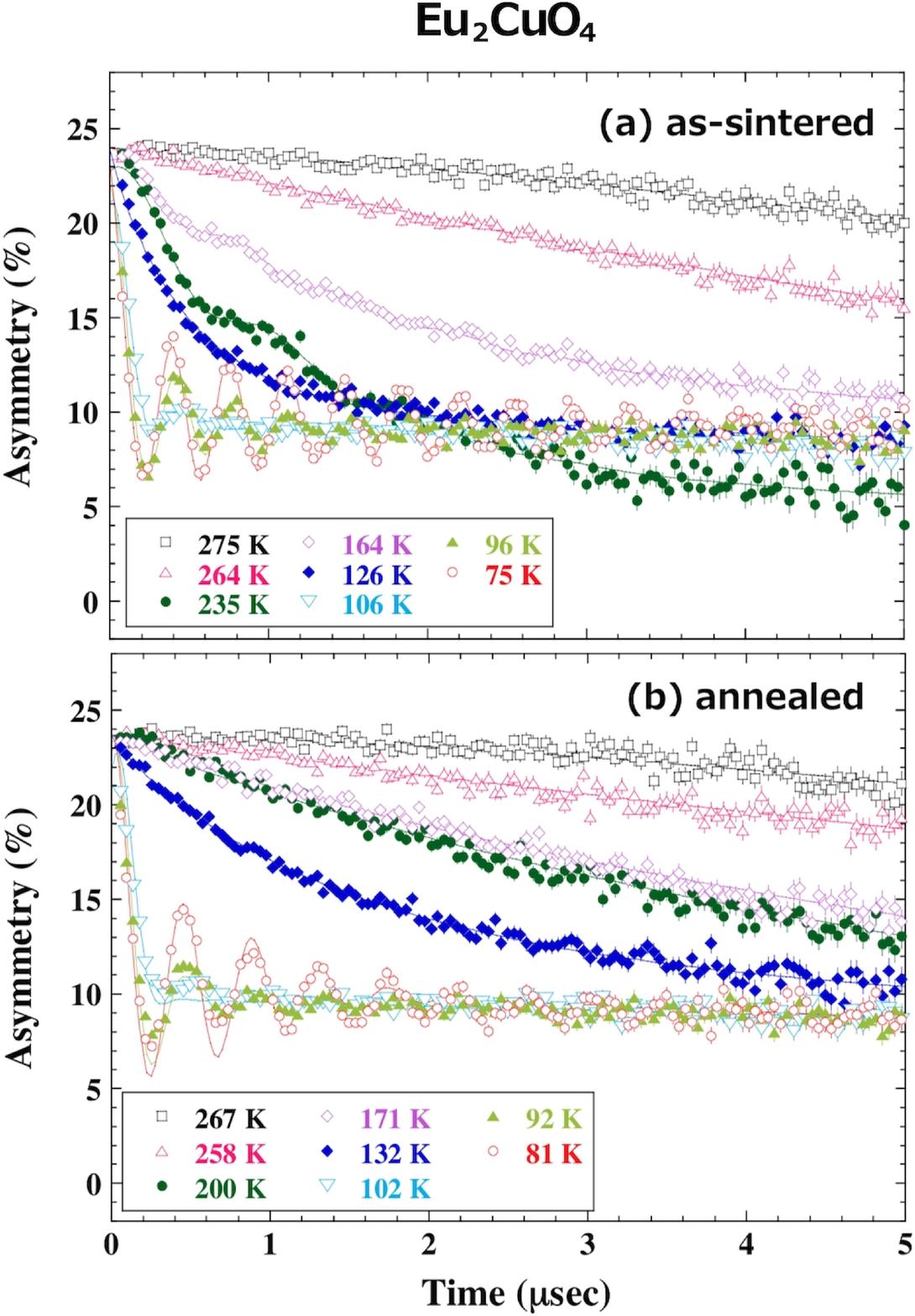}
	\caption{Time spectra of zero-field $\mu$SR for (a) as-sintered and (b) annealed Eu$_2$CuO$_4$. Solid curves indicate the results of least-squares fitting based on Eq. (\ref{eq1})--(\ref{eq3}). All the data were acquired at MLF.}
	\label{ZFspectraECO}
\end{figure}

Figure \ref{ZFspectraECO}(a) shows the time spectra of ZF $\mu$SR for AS ECO at representative temperatures. A Gaussian-type slow decay due to the nuclear dipole fields is observed at 275 K. Upon cooling to 264 K, an exponential-type spectrum appears, indicating a slowdown of electron spin fluctuations. With further cooling, the spectra show faster decay at 235 K and, subsequently, another phase of slower decay at 164 K. This trend is not a typical ordering sequence; that is, the spin fluctuations slow down monotonically toward the freezing temperature. Furthermore, in the spectra at 164 K and 235 K, an oscillation component with a small amplitude is superimposed on the dominant exponential-type decay. Therefore, a partially ordered spin state coexists with a fluctuating spin state and/or spin-glass-like state at these temperatures. The small oscillation component disappears at 126 K, resulting in a simple exponential-type decay spectrum. A well-defined oscillation spectrum appears at even lower temperatures. Thus, the true magnetic order occurs at a much lower temperature than the previously reported temperature of 265 K~\cite{Chattopadhyay1994}, and the ordering temperature is comparable to the $T_\mathrm{N2}$ of 110 K for $T^{\prime}$-type LCO~\cite{Hord2010}. Here, we emphasize that the existence of the small oscillation component and the detailed thermal evolution of magnetism on the CuO$_2$ plane were revealed by the high-quality data with fine temperature steps. 

Figure \ref{ZFspectraECO}(b) shows the ZF $\mu$SR spectra of AN ECO. 
Similar to the results for AS ECO, exponential-type depolarization appears at $T_\mathrm{N1} = 250$ K upon cooling (see the spectra at 200, 171, and 132 K). The spectrum at 171 K shows a slower time decay than that at 200 K. It transforms into a faster decay spectrum again at 132 K, following which oscillation spectra appear below $T_\mathrm{N2} = 110$ K. Thus, the sequential development of magnetism with a low $T_\mathrm{N2}$ is the same in AS and AN ECO. One of the qualitative differences between the spectra of AS and AN ECO is the small oscillation component around 200 K, which is absent in the AS ECO spectra. Another difference is the considerably weaker depolarization in AN ECO around this temperature. These results indicate that the magnetism at higher temperatures is varied considerably by annealing.

For quantitative evaluation of the thermal evolution of magnetism, we analyzed the time spectra using the following functions: 
	\begin{eqnarray}
\begin{split}
	A(t) = A_0\{
	f_\mathrm{se}e^{-(\lambda _\mathrm{se}t)^\beta}
	+ f_\mathrm{osc} e^{-\lambda _\mathrm{d1}t} \cos (\omega _1t + \phi _1)\\
	+ (1 - f_\mathrm{se} - f_\mathrm{osc}) e^{-\lambda _\mathrm{s}t}\} +A_\mathrm{BG}   \hspace{5mm}
	  (T > T_\mathrm{N2}),
	\label{eq1}
\end{split}
	\end{eqnarray}
	\begin{eqnarray}
\begin{split}
	A(t) = A_0 \{f_\mathrm{osc} G_\mathrm{T}(t) 
	+ (1-f_\mathrm{osc})e^{-\lambda _st}\}
	+ A_\mathrm{BG} \\
	  (T \leq T_\mathrm{N2}), 
	\label{eq2}
	\end{split}
	\end{eqnarray}
where 
	\begin{eqnarray}
\begin{split}
	G_\mathrm{T}(t) =
	r_1 e^{-\lambda _\mathrm{d1}t} \cos (\omega _1t + \phi _1)
	+ (1 - r_1) e^{-\lambda _\mathrm{d2}t}. 
	\label{eq3}
\end{split}
	\end{eqnarray}
The first (third) and second terms in Eq. \ref{eq1} represent the fast (slow) decay and the small oscillation components, respectively. Similarly, the first and second terms in Eq. \ref{eq2} represent the oscillation and decay components, respectively. 
$A_0$ is the initial asymmetry at $t = 0$, and $A_\mathrm{BG}$ is the time-independent background. Both $A_0$ and $A_\mathrm{BG}$ were fixed throughout the analysis. $f_\mathrm{se}$ ($f_\mathrm{osc}$) is the fraction of the stretched exponential (oscillation) component. $\lambda_\mathrm{se}$ and $\beta$ are the relaxation rate and power of the stretched exponential component, respectively. $\lambda _\mathrm{d1}$ and $\lambda _\mathrm{d2}$ are the damping rates of the oscillations, and $r_1$ is the ratio of the damping components. $\omega _1$ and $\phi _1$ are the frequency and phase of the oscillation, respectively.

\begin{figure}[t]
	\includegraphics[width=0.66\linewidth]{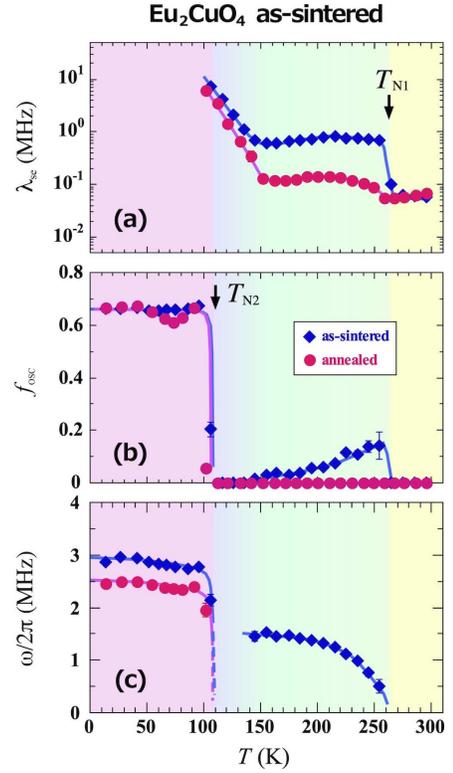}
	\caption{Temperature dependencies of (a) the depolarization rate of the stretched exponential term $\lambda _\mathrm{se}$, (b) the asymmetry fraction of the oscillation component $f_\mathrm{osc}$, and (c) the oscillation frequency $\omega_{1}$ for as-sintered and annealed Eu$_2$CuO$_4$.
	The spectra, acquired at MLF and RAL, were analyzed using Eq. (\ref{eq1})--(\ref{eq3}) to obtain the parameters. 
	}
	\label{paramECO}
\end{figure}

\begin{figure}[t]
	\includegraphics[width=0.68\linewidth]{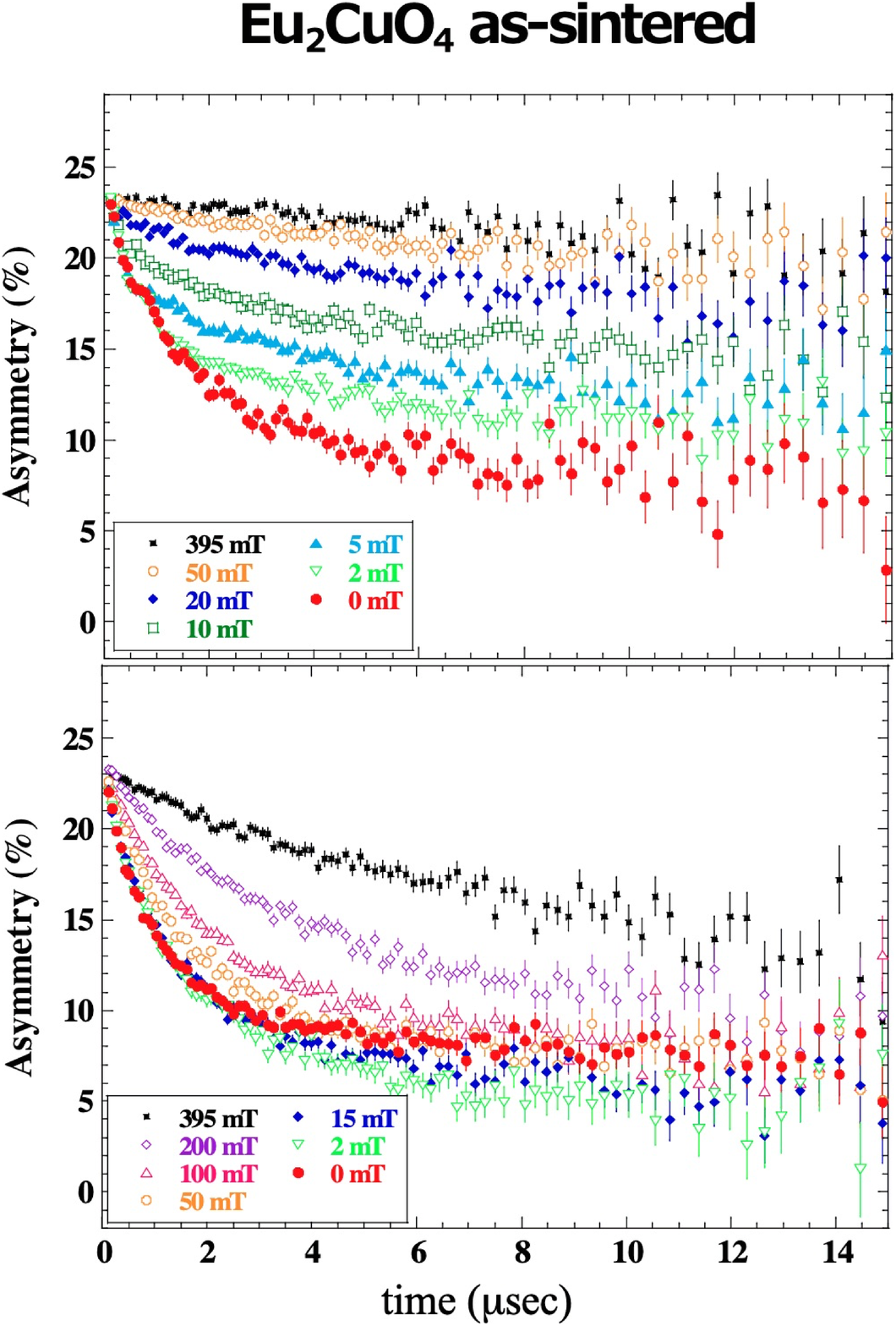}
	\caption{Field dependencies of $\mu$SR time spectra in longitudinal magnetic fields for as-sintered Eu$_2$CuO$_4$ at (a) 187 and (b) 138 K. 
	All the data for Eu$_2$CuO$_4$ were acquired at MLF.}
	\label{LFspectraECO}
\end{figure}

As shown in Fig. \ref{paramECO}(a), AS and AN ECO exhibit similar temperature dependencies of $\lambda _\mathrm{se}$, i.e., with cooling, $\lambda _\mathrm{se}$ increases sharply at $T_\mathrm{N1}$, maintains a constant value until $\sim$150 K, and increases further below this temperature. However, there are apparent differences between the two samples; in AS ECO, the magnetism around $T_\mathrm{N1}$ develops in a rather broader temperature range, and the magnitude of $\lambda _\mathrm{se}$ below $T_\mathrm{N1}$ is much smaller. These results indicate the enhancement of spin fluctuations by annealing. 
Upon cooling, the small oscillation component in AS ECO disappears with a reduction in $f_\mathrm{osc}$, as shown in Fig. \ref{paramECO}(b). This disappearance proceeds concomitantly with the development of $\lambda _\mathrm{se}$ toward $T_\mathrm{N2}$, suggesting an interrelation between the partially ordered and fluctuating spin states. In both AS and AN ECO, bulk long-range order is observed below $T_\mathrm{N2}$ of 110 K, which is comparable to $T_\mathrm{N2}$ for $T^{\prime}$-type LCO~\cite{Hord2010}. Therefore, the static order with a low $T_\mathrm{N2}$ is a robust feature of $T^{\prime}$-type $R$CO. Furthermore, $\omega_1$ does not exist for 110 K $\lesssim T \lesssim$ 150 K owing to the disappearance of the small oscillation component. The extrapolation of $\omega_1$ at $150 \:\mathrm{K} < T < T_\mathrm{N1}$ toward lower temperatures does not seem to connect smoothly with the values of $\omega_1$ below $T_\mathrm{N2}$. This discontinuity and the disappearance of the small oscillation component immediately above $T_\mathrm{N2}$ suggest that the partial order at high temperatures is not a direct precursor of the static magnetic order below $T_\mathrm{N2}$. 

\subsection{Longitudinal-field $\mu$SR measurements}


As mentioned in Section \ref{ZF}, we found an anomalous ordering sequence above $T_{\rm N2}$ in AS ECO. 
Therefore, we performed LF $\mu$SR measurements to obtain further information on the spin states above $T_\mathrm{N2}$. 
Figures \ref{LFspectraECO}(a) and (b) show the field dependencies of the $\mu$SR spectra for AS ECO at 187 and 138 K, respectively. At 187 K, where the small oscillation component was observed,  the spectra in the longer time region exhibited a parallel upward shift under the magnetic fields. 
This shift can be observed in the decoupling process of the interaction between muon spins and spontaneous local magnetic fields at the muon sites, providing the magnitude of the spontaneous magnetic fields. In AS ECO at 187 K, an external field of 20 mT is sufficient to shift the overall spectra over the entire time range, suggesting that the static magnetic fields at the muon sites are less than 20 mT in magnitude. Therefore, the partially and weakly ordered spin state exists at 187 K. At 138 K,  where the small oscillation component was absent, the spectra slightly changed with an external field of 50 mT, and relaxation remained at 395 mT, as shown in Fig. \ref{LFspectraECO}(b). These results indicate that a dynamical nature indeed develops at a lower temperature of 138 K. 
This anomalous behavior is discussed in relation to the inhomogeneous spin state in Sec. \ref{Discussion3}. 


\section{Discussion}\label{Discussion}

\subsection{Common feature of magnetism}\label{Discussion1}

The present $\mu$SR measurements demonstrate that a uniform static magnetic order exists only below $T_\mathrm{N2}$ (= 110 K) and a fluctuating spin state exists for $T_\mathrm{N2} < T < T_\mathrm{N1}$ in AS and AN ECO. These results are consistent with previous $\mu$SR studies for $T^{\prime}$-type LCO~\cite{Luke1990, Hord2010} and contradictory to the absence of a fluctuating state with a higher N{\'e}el temperature ($T_\mathrm{N2} \approx T_\mathrm{N1}$ = 325 K) in $T$-type LCO~\cite{Uemura1989}. 
Therefore, the sequential ordering process with a lower $T_\mathrm{N2}$ is a common feature of magnetism of $T^{\prime}$-type $R$CO. 
The variation of the magnetic order in the parent $R$CO with different structures most likely originates from differences in oxygen coordination. 
The spin direction can be perturbed by the contribution of the Cu 3$d_{3z^2-r^2}$ orbital to the spin correlations, causing the magnetic order to be more stabilized in $T$-type $R$CO having apical oxygen atoms. 

The fact that AS $T^{\prime}$-type ECO ($a$ = 3.9049 {\AA}) and LCO ($a$ = 4.0102 {\AA}~\cite{Hord2010}) with different lattice constants $a$ have similar magnetic behaviors is in contrast to the variation of SC properties in AN $R$CO depending on the lattice constant $a$. In $T^{\prime}$-type $R_{2-x}$Ce$_x$CuO$_4$ with a larger $a$, the critical Ce concentration for the appearance of the SC phase is smaller\cite{Naito2002, Fujita2003, Krockenberger2008, Krockenberger2014} and the optimal SC transition temperature is higher\cite{Naito2002, Maple1990}. Thus, the appearance of non-doped superconductivity is not directly related to the lower $T_\mathrm{N2}$. 

\begin{figure}[t]
	\includegraphics[width=70mm]{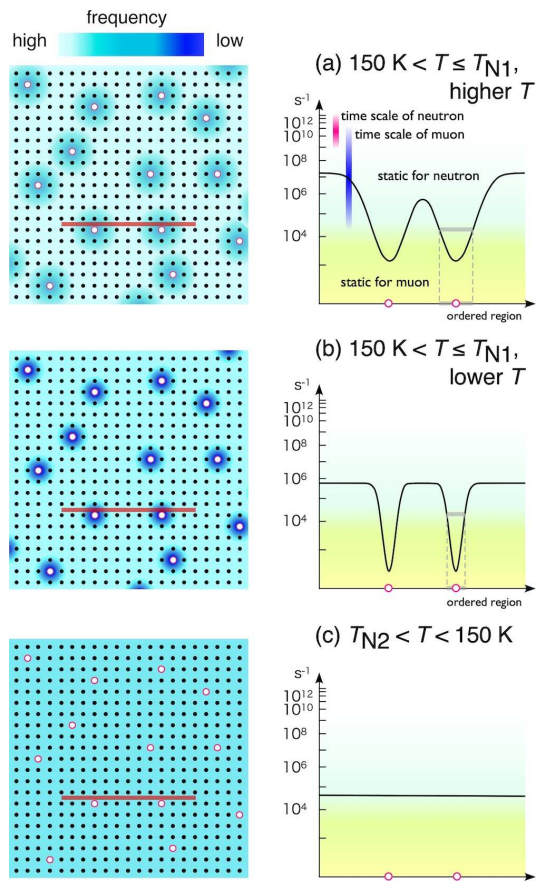}
	\caption{Schematics of the spin states on the CuO$_2$ plane in the as-sintered sample for (a) higher temperatures in $150 \:\mathrm{K} < T < T_\mathrm{N1}$, (b) lower temperatures in $150 \:\mathrm{K} < T < T_\mathrm{N1}$, and (c) $T_\mathrm{N2} < T < 150$ K. The open circles represent the chemical defect that can function as pinning centers of spin fluctuations. The figures show the case with $\sim$3$\%$ defect. The right column shows the corresponding position dependence of the frequency of spin fluctuations along the red line. The typical time scales of neutron scattering and muon spin rotation/relaxation measurements are indicated by color bars in the top figure.}
	\label{schematicview}
\end{figure}

We note that the neutron diffraction measurements of $T^{\prime}$-type $R$CO showed magnetic Bragg peaks below approximately $T_\mathrm{N1}$ but no evidence of an anomaly around $T_\mathrm{N2}$~\cite{Chattopadhyay1994}. The different ordering temperatures determined using the $\mu$SR and neutron diffraction measurements originate from the different time scales to observe the spin dynamics, which are typically $\sim$10$^{-4}$ s to $\sim$10$^{-11}$ s and $\sim$10$^{-9}$ s to $\sim$10$^{-14}$ s, respectively. Therefore, a fluctuating spin state with a time scale of $\sim$10$^{-4}$ to $\sim$10$^{-9}$ s would be realized for $T_\mathrm{N2} < T \leq T_\mathrm{N1}$. (See Fig. \ref{schematicview}.)

\subsection{Effect of annealing on long-range magnetic order}\label{Discussion2}

The negligible annealing effect on $T_\mathrm{N2}$ in $T^{\prime}$-type ECO appears to be different from the annealing-induced suppression of AFM order reported for $T^{\prime}$-type LECO. As mentioned in the Introduction, reduction annealing plays two roles: electron doping\cite{Asano2018} and the removal of chemical defect~\cite{Radaelli1994c, Schultz1996, Kang2007}. Because the $\delta$ values for the present ECO and LECO samples are comparable~\cite{Asano2020, Takagi1989, Armitage2010}, the overall structural change due to annealing is expected to be the same for both samples. 

Recent x-ray absorption near-edge structure measurements revealed a Mott insulating state in both AS LECO and NCO but a significant increase of electron density due to annealing ($n_{\rm AN}$) in LECO (0.4 electrons per Cu), which is much larger than that in NCO (0.05 electrons per Cu)~\cite{Asano2020}. On this basis, a distinct electron-doping process was proposed: the self-doping of electrons and holes associated with the collapse of the charge-transfer gap for the former with a smaller gap size ($\Delta_{\rm CT}$) and electron doping into the upper Hubbard band for the latter with a larger $\Delta_{\rm CT}$. According to this scenario, the present ECO sample, which has a relatively large $\Delta_{\rm CT}$~\cite{Arima1991, Braden2005}, follows the $n_{\rm AN}=2\delta$ relation~\cite{Asano2018}, as is the case for NCO. 
Therefore, the distinct effects of annealing on the long-range magnetic order in ECO and LECO~\cite{Adachi2016} can be interpreted as due to different induced carrier densities. The suppression of the ordered state in LECO is attributed to the generated charges with a large $n_{\rm AN}$ value, while the negligible variation in $T_\mathrm{N2}$ and the small reduction of $\omega_{\rm 1}$ in ECO are likely to have originated from electron doping with a low $n_{\rm AN}$ value (= 2$\delta$ $\sim$0.04 electrons per Cu).

\begin{figure}[t]
	\includegraphics[width=0.82\linewidth]{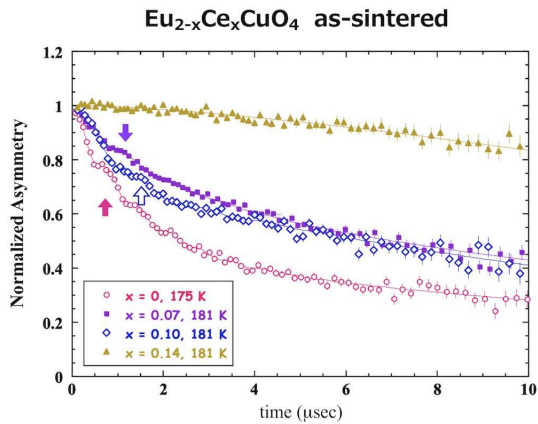}
	\caption{Normalized time spectra of zero-field $\mu$SR for as-sintered Eu$_{2-x}$Ce$_x$CuO$_4$ with $x$ = 0, 0.07, 0.10, and 0.14 measured at approximately 180 K. All the data were acquired at MLF. The arrows indicate the small oscillation component. The solid lines indicate the best-fitting results based on Eq. (\ref{eq1}).}
	\label{LFspectraECCO}
\end{figure}

\subsection{Partially ordered spin state}\label{Discussion3}

Next, we discuss the partially ordered spin state in AS ECO. Because this state disappears with annealing, its origin is argued from the viewpoint of the annealing effect. To exclude the possibility of electron doping aspect of annealing treatment, we performed additional ZF $\mu$SR measurements on AS Eu$_{2-x}$Ce$_x$CuO$_4$ (ECCO) with $x$ = 0.07, 0.10, and 0.14. If the short-range magnetic order is related to the doping level, the small rotation component should vanish in AS ECCO with $x$ $\geq$ 0.04, the electron density in which is greater than that in AN ECO. Figure \ref{LFspectraECCO} shows normalized time spectra at $\sim$180 K. As indicated by arrows, the small oscillation component can be observed on the dominant exponential-type decay component for $x$ $\leq$ 0.10. Therefore, in contrast to long-range magnetic order, which is varied by the electron-doping effects of the annealing process, the partial order for 150 $<$ T $<$ $T_\mathrm{N1}$ is an intrinsic property of AS compounds independent of the doping level. Note that in $x$ = 0.14, no evidence of oscillation was detected at 181 K, because this temperature is higher than $T_\mathrm{N1}$. 

To understand the origin of the partial order, the appearance of oscillation immediately below $T_\mathrm{N1}$ is noteworthy because it indicates a close connection between the partially ordered state and the dominant fluctuating spin state. In the slowly fluctuating spin state, chemical defect  could become pinning centers for the fluctuating spins and induce local magnetic order~\cite{Adachi2013}. 
The structural feature in AS samples containing chemical defects is qualitatively consistent with the appearance of the small oscillation component below $T_\mathrm{N1}$. Reduction annealing can remove such defects\cite{Radaelli1994c, Schultz1996, Kang2007}, resulting in the suppression of local magnetism and the enhancement of the dynamical spin nature. In AN ECO, the non-monotonic evolution of the muon spin relaxation suggests that a tiny amount of defects remains.

Finally, we discuss the disappearance of the small oscillation component and the anomalous enhancement in the dynamical spin nature of AS ECO for $T_\mathrm{N2} < T < 150$ K with cooling. 
When the correlation length of the pristine fluctuating spin state develops toward static order, the partially and weakly ordered region contributes to the dominant state owing to the increase of magnetic energy. Consequently, the partial order could disappear immediately above $T_\mathrm{N2}$, where the correlation length of the fluctuating state diverges~\cite{Endoh1989, Birgeneau1999, Motoyama2007}. Based on the present results, this thermal evolution of the magnetic behavior is schematically drawn in Fig. \ref{schematicview}. 
The figure depicts the case in which $\sim$3$\%$ of pinning centers (open circle) exists on the CuO$_2$ plane. At high temperatures below $T_\mathrm{N1}$, static order is locally induced around the pinning center owing to the slowing down of spin fluctuations in the time scale of the $\mu$SR measurement. The ordered region shrinks upon cooling and vanishes immediately above $T_\mathrm{N2}$. The figures in the right column show the corresponding position dependence of the frequency of spin fluctuations along the red line near the pinning centers. All these states are observed as static order in the neutron scattering measurements.

\section{Summary}

We have performed ZF and LF $\mu$SR measurements on $T^{\prime}$-type ECO to study the common features of magnetism in the parent compounds of electron-doped superconductors with the $T^{\prime}$-structure. The effects of reduction annealing on the magnetism were also investigated. 
We clarified the development of spin correlations below a $T_\mathrm{N1}$ of 265 K and the formation of true static magnetic order in both AS and AN ECO below a $T_\mathrm{N2}$ of 110 K. $T_\mathrm{N2}$ is comparable to the ordering temperature in $T^{\prime}$-type LCO and much lower than that in $T$-type LCO. Thus, the sequential ordering process and low $T_\mathrm{N2}$ are common features of magnetism in non-superconducting $T^{\prime}$-type $R$CO. 
Furthermore, we found in AS ECO the coexistence of a dominant fluctuating spin state and partially ordered spin state for $\sim$$150 \:\mathrm{K} < T \leq T_\mathrm{N1}$. The partially ordered spin state is absent in AN ECO, indicating that this state originates from chemical defect in the AS sample.


\section*{Acknowledgments}

The $\mu$SR measurements at the Materials and Life Science Experimental Facility of J-PARC (Proposals No. 2013A0084, No. 2014A0205, No. 2015MP001, and No. 2017B0270) and at the RIKEN-RAL Muon Facility in the Rutherford Appleton Laboratory (Proposal No. RB1670580) were performed under user programs. 
We thank the staff at J-PARC and RIKEN-RAL for their technical support during the experiments. We are also thankful for helpful discussions with T. Adachi, Y. Ikeda, Y. Koike, K. M. Kojima, Y. Nambu, and K. Yamada. This work was supported by MEXT KAKENHI under Grants No. 15K17689 and No. 16H02125 as well as the IMSS Multi-probe Research Grant Program.


\bibliographystyle{apsrev4-1}
\bibliography{Eu2CuO4}

\end{document}